\documentclass[12pt]{article}
                      
\usepackage[round]{natbib}                                                               
\RequirePackage[OT1]{fontenc}
\usepackage{amsthm,amsmath,bm, bbm}                                                      
\usepackage[a4paper]{geometry}                                                           
\RequirePackage[colorlinks, citecolor=blue,urlcolor=black, linkcolor=black]{hyperref}    
\RequirePackage{pifont, amssymb}                                                         
\RequirePackage{enumerate}
\usepackage{array,graphicx}
\usepackage{xifthen}
\usepackage{xparse}
\usepackage{todonotes}

\RequirePackage{tikz, tkz-tab, pgfplots}                                                 
\usetikzlibrary{trees, calc, positioning, arrows.meta,positioning,quotes, shapes}        
\RequirePackage{rotating, pdflscape} 
\RequirePackage{multirow, array}      
\newcolumntype{L}[1]{>{\raggedright\let\newline\\\arraybackslash\hspace{0pt}}m{#1}}

\newcolumntype{C}[1]{>{\centering\let\newline\\\arraybackslash\hspace{0pt}}m{#1}}
\newcolumntype{R}[1]{>{\raggedleft\let\newline\\\arraybackslash\hspace{0pt}}m{#1}}
\newcommand{\E}[1]{\mathbb{E}\left(#1\right)}

\newcommand{\fY}[2]{\ifthenelse{\isempty{#2}}{\gamma_{#1}}{\gamma_{#1,#2}}}                      
\newcommand{\fe}[2]{\ifthenelse{\isempty{#2}}{\varphi_{#1}}{\varphi_{#1,#2}}}                    
\newcommand{\fX}{\beta}                                                                          

\newcommand{\rZ}[2]{\ifthenelse{\isempty{#2}}{c_{#1}}{c_{#1,#2}}}                                
\newcommand{\rforecast}[2]{\ifthenelse{\isempty{#2}}{\ell_{0,#1}}{\ell_{0,#1,#2}}}               
\newcommand{\rYtilde}[3]{\ifx\\#3\\%
	{\ifx\\#2\\%
		\tilde{{g}}_{#1}     
		\else
		\tilde{{g}}_{#1,#2}  
		\fi}
	\else
	\tilde{{g}}_{#1,#2,#3}   
	\fi}

\newcommand{\rY}[3]{\ifx\\#3\\%
	{\ifx\\#2\\%
		g_{#1}     
		\else
		g_{#1,#2}  
		\fi}
	\else
	g_{#1,#2,#3}   
	\fi}

\newcommand{\re}[3]{\ifx\\#3\\%
	{\ifx\\#2\\%
		f_{#1}     
		\else
		f_{#1,#2}  
		\fi}
	\else
	f_{#1,#2,#3}   
	\fi}

\newcommand{\YY}[2]{\ifthenelse{\isempty{#2}}{Y_{#1}}{Y_{#1, #2}}}                           
\newcommand{\YYhat}[2]{\ifthenelse{\isempty{#2}}{\hat{Y}_{#1}}{\hat{Y}_{#1, #2}}}    
\newcommand{\ee}[2]{\ifthenelse{\isempty{#2}}{e_{#1}}{e_{#1, #2}}}
\newcommand{\design}[3]{\ifthenelse{\isempty{#3}}{#1_{#2}}{#1_{#2, #3}}}                                                         
\newcommand{\ww}[2]{\ifthenelse{\isempty{#2}}{w_{#1}}{w_{#1, #2}}}

\begin{document}

\vfill
\begin{center}
	\Large
	\bfseries
	Random autoregressive models:\\ A structured overview\\
	\vspace{2cm}
	\normalsize
	\mdseries
Marta Regis (m.regis@tue.nl)$^*$\footnote{ORCiD:0000-0003-4306-8673;  email: \url{m.regis@tue.nl}}\\
Paulo Serra (p.j.de.andradeserra@vu.nl)$^{**}$\\
Edwin R. van den Heuvel (e.r.v.d.heuvel@tue.nl)$^*$\\
\vspace{1cm}
\footnotesize
$^*$Eindhoven University of Technology, Department of Mathematics and Computer Science\\ P.O. Box 513, 5600 MB Eindhoven, The Netherlands.\\
$^{**}$Vrije Universiteit Amsterdam, Department of Mathematics \\ De Boelelaan 1111, 1081 HV Amsterdam, The Netherlands.\\
\vspace{8cm}
\normalsize
\end{center}

\newpage

\begin{abstract}
Models characterized by autoregressive structure and random coefficients are powerful tools for the analysis of high-frequency, high-dimensional and
volatile time series. The available literature on such models is broad, but
also sectorial, overlapping, and confusing. Most models focus on one property of the data, while much can be gained by combining the strength of
various models and their sources of heterogeneity.

We present a structured overview of the literature on autoregressive models with random coefficients.
We describe hierarchy and analogies among models, and for each we systematically list properties, estimation methods, tests, software packages and typical applications.
\end{abstract}

\bfseries Keywords \mdseries (Generalized) Random coefficient autoregressive models;  (Generalized) Autoregressive conditional heteroskedasticity models; 
Autoregressive panel data models; Time-series-cross-section models; Random coefficient panel models

\newpage
\section{Introduction}\label{sec:Introduction}
(Macro)economic studies are often characterized by a large number of units observed a relatively small number of times (e.g.~population and income shares for a large number of countries).
On the other hand, data-sets in financial studies consist of high-frequency, volatile single-unit time series (e.g.~daily trading volume of IBM stock and annual change rates of world GDP).
With the increased capability of storing information, data-sets are growing in all dimensions with economic variables being monitored at higher frequency or for longer time, and financial time series being collected for multiple units. This makes the distinction between the two types of data more nuanced, and requires sophisticated statistical methods to handle multiple sources of heterogeneity.																																																								

Models have been proposed in the literature of the economic and financial fields to address pooling of short time series  and handling volatility respectively. 
Although the two purposes seem very different, these models share, in most cases, an autoregressive structure and the inclusion of random coefficients.
As a result these references constitute a very large literature, that is hard to navigate.
The two fields are hardly connected and the terminology is not unified, resulting in the same model being defined differently or different mathematical models being referred to in the same way.
For instance, the random coefficient autoregressive (RCA) model of~\cite{Nicholls1982} used in financial applications and the random coefficient autoregressive model of~\cite{Liu1980} considered in biological studies share the same name but are different.
The first model aims at capturing volatility in time for a single time series, while the goal of the second is to represent heterogeneity among units under the same modeling assumptions.
On the other hand, the RCA of~\cite{Liu1980} and the autoregressive panel data model of~\cite{Nandram1997} are in fact the same model, although they do not share the same name.
Something similar happens for estimation methods, that are often defined multiple times with different names, while having identical properties. 
In the introduction to their work, \cite{Chandra2001} emphasize that various approaches that developed independently in the fields of time series and panel data are often very similar. For instance, they observe that ``estimating functions and generalized method of moments are essentially the same''. On the other hand, they refer to the model in hand as RCA, while it is in fact its generalized version.
Some authors have pointed out the risk of confusion.
For example~\cite{Andvel1976} highlights the different sampling of the random coefficients in the RCA model of~\cite{Nicholls1982} from that of~\cite{Liu1980}, while~\cite{Hsiao2014} compares models for panel data and RCA models, commenting on certain peculiarities.
However, these resources are insufficient to guide researchers and practitioners across such an extensive literature.
Furthermore, methods have also been investigated in theoretical statistical and econometric studies, and new developments and applications are now foreseen in medicine, engineering, psychology, sociology and politics.
Also in this case, cross-referencing is not always guaranteed, contributing to additional duplication, confusion and fragmentation.

With the present work we aim at giving a comprehensive, structured overview of existing autoregressive models with random coefficients.
We fill the gap between two traditionally independent families of models that capture heterogeneity across units and heterogeneity across time respectively (and can be linked to the economic and financial fields accordingly). This creates a solid basis for further developments in handling multiple sources of heterogeneity, in line with recent publications~\citep{Horvath2016}.
For each model we illustrate advantages and shortcomings, properties, estimation methods, tests, software packages and typical applications. 
We also investigate mutual analogies and differences.

To accomplish this task, we define the random autoregressive moving average (RARMA), a unifying structure that captures all other models as particular cases.
The RARMA is introduced here for classification purposes, to make the exposition clear.
In fact, it provides a unified language to better connect and contrast existing models, and by using mathematical rigorousness it shows how well-known random coefficient models can be obtained from the general equation, by imposing certain restrictions.
We support this mathematical overview with a graph to visually represent the hierarchy and a table to have all models and their main characteristics at a glance.
This way, the explanation is visually and mathematically clear, and thus more easily accessible.

Our overview guides the reader through the terminology used, the models existing in the literature, and the choice of the most appropriate structure (with relevant inference, tests and software packages) for specific types of data.
This way, the RARMA structure creates a solid basis for further exploration and development of heterogeneous time series models for complex data.

The paper is organized as follows: Section~\ref{sec:UnifiedOutline} presents a consolidated outline, presenting the RARMA and the model families existing in the literature.
The definition of the RARMA structure is provided in Section~\ref{sec:RAR} together with its assumptions.
Then, following the structure given in Section~\ref{sec:UnifiedOutline}, we move down the hierarchy and focus on the two traditionally independent families of models: models for the heterogeneity in time (Section~\ref{sec:TimeHeterogen}), and models for the heterogeneity across units (Section~\ref{sec:UnitHeterogen}).
We state how to obtain each model equation from the general one, gather related contributions from the literature, and discuss the main properties and estimation methods for these models.
Section~\ref{sec:TimeUnitHeterogen} is dedicated to recent model structures, capable of addressing both forms of heterogeneity.
Finally, we end the work with a discussion in Section~\ref{sec:Discussion}.

\section{Unified outline}\label{sec:UnifiedOutline}
The literature that we consider here is broad and heterogeneous, and the terminology often changes according to the field of application.
This commonly causes some disorientation when first coming into contact with this literature, sometimes needlessly isolating entire fields.
Authors often propose already existing definitions unaware of previous uses, or refer to the same object in different ways.
In this section we provide an overview of the terms which have been used in the literature to describe both the data and the models.
A complete list of all the definitions introduced in the past is beyond the scope of this work.
However, the information provided here will considerably ease the reader's navigation of the existing literature.
We also introduce the terminology that we use in the remainder of this review.
Since a broad set of definitions already exists, we avoid introducing new terms unless strictly necessary.
We stick to the existing nomenclature, even if it is somewhat unclear, and only modify it if the same term is used for two different objects, or make additions whenever we define a new object.
The families of models are then presented with the help of a graph, making the hierarchy clear at a first glance.
In the remainder of the overview, the models in the graph are introduced one by one and discussed. 

\subsection{Terminology for data structures}
Data is characterized by three dimensions:
the number of units/subjects $n$,
the number of outcomes $m$, and
the number of samples $T$.
The relative sizes of these dimensions can vary significantly, depending on the research study and application.

The number of observed units $n$ is typically large in economic and clinical studies.
Here the focus is on pooling information from the $n$ units, since the number of repeats $T$ per unit is often limited.
For this type of data, a variety of terms has been introduced such as {\em panel data}~\citep{Hsiao2014}, {\em panel of time series}~\citep{Franses2006}, {\em time series panel data}~\citep{Nandram1997} and {\em time-series-cross-section data}~\citep{Beck1998}.
All of these definitions aim to underline the extension from traditional single unit time series to multiple units.
Note however that in social sciences and medical research the term {\em cross-sectional} traditionally refers to the analysis of the population at a fixed time point -- and that this clashes with the meaning that this term takes in the context of time series.
When considering the dimension $m$ of the response variable, the terminology normally used is {\itshape uni-dimensional or univariate} for $m=1$ and {\itshape multidimensional or multivariate} for $m>1$.
In some cases, $m=2$ is distinguished from the multivariate case with the term {\itshape two-dimensional or bivariate}. 
In the literature considered in the present work, $m$-dimensional responses are rarely addressed.

Finally, the third dimension is the number of observations $T$.
Data with a high number of repeats on a single unit is normally the object of interest in finance, and is referred to as {\em time series data}.
In these applications in fact, the number of units is often not mentioned and taken equal to one.
The terms {\em intensive/high-frequency longitudinal\/} also refer to repeated observations, and normally the distinction between {\em longitudinal\/} and {\em time series\/} data is either in the number of repeats, which is much larger for the latter, or in the number of observed units, which is larger for longitudinal data.

With the increasing amounts of data being collected, the three dimensions are often all large at the same time, making the choice of the terminology not easy.
Thus we propose the use of a very simple nomenclature, addressing the first two dimensions by {\em single/multiple-units} and {\em univariate/multivariate} respectively.
Then, since all of the data that we consider have more than two repeats, we will call it {\em time series data}.
In this way, we can have for example a {\em multiple-units univariate time series}, a {\em single-unit multivariate time series} and so on.
Furthermore, we propose a shortened version through which one can also specify precisely the magnitude of each dimension, namely {\em$(n, m,T)-$data}.

\subsection{Terminology for models}
Models for the analysis of temporal data are known in econometrics and statistics as {\em time series models\/}.
In particular, we focus here on {\em autoregressive models\/}, i.e.~models regressing the outcomes on previous values of the same time series.
In the economic literature, models with this structure are also defined as {\em dynamic panel data models} to be distinguished from {\em static panel data models} which do not contain lagged dependent variables.
Models used for repeated measurements recorded at lower frequency on multiple units are referred to as {\em longitudinal models\/} in statistics. 

All models treated in the overview are in Table~\ref{Table:Acronyms}, together with their acronyms. 
In the same table we explain how the definitions used here differ from the ones used in the literature, what the random effects are, and how they are sampled.
Furthermore, we specify in which section the model is treated.
We distinguish two families of models (the blue dashed cluster on the left hand side, and red dotted box on the right hand side of Figure~\ref{fig:structure} respectively) based on how the random terms are sampled:
{\em models for heterogeneity in time} have the coefficients that are stochastic processes in time, while {\em models for heterogeneity across units} sample their coefficients from a certain population described by a probability distribution.

These two families do not usually overlap in the literature.
The first is typical in finance and developed around the RCA model of~\cite{Nicholls1982}.
The second has grown in parallel to accommodate economic variables, has expanded theoretically in econometrics and statistics, and towards biological and psychological applications.
Also the asymptotics vary substantially depending on the research question and type of data in hand:
asymptotic behavior of estimators is studied with respect to the number of observations $T\longrightarrow \infty$ and fixed $n$ when the focus is on the heterogeneity in time or for increasing number of units $n\longrightarrow \infty$ and fixed amount of repeats when the focus is on pooling many units.
Recent models addressing both heterogeneity in time and across units (random coefficient panel data, and dynamic factor models) lie in the overlap of both families.
For this and few other models, the asymptotic behavior is studied both with respect to time and unit.

The acronyms in the nodes of Figure~\ref{fig:structure} correspond to those listed in Table~\ref{Table:Acronyms}.
The arrows linking two nodes point from the more general model to a sub-model that has given rise at least to an independent publication, while the dashed lines highlight looser connections and similarities.
We introduce the random autoregressive moving average (RARMA) model to gather all the considered models under a single structure.
Since the literature on the topic is extremely vast, we have limited our overview to models fulfilling the following inclusion criteria:
\begin{enumerate}
	\item autoregressive structure,
	\item at least one parameter is randomly sampled from a distribution and
	\item discrete-time models.
\end{enumerate}
The gray nodes mark the models that are characterized by the three properties above.
For completeness we include also some structures which do not satisfy all three properties, but are closely related to the main topic of the overview.
These models are represented by white nodes in Figure~\ref{fig:structure}, and shortly discussed in the remainder.\\ 

\newpage
\begin{sidewaystable}[http!]
	\footnotesize
	\centering
	\renewcommand{\arraystretch}{1.1}
	\begin{tabular}{cC{7cm}C{2cm}cC{2.5cm}c}
		\hline
		Acronym & Full model name                                               & Changes & Section         & Random term                   & Heterogeneity \\
		\hline\hline
		ARCH    & Autoregressive conditional heteroscedasticity                 & -       & 4.3  & -                             & time \\ 
		\hline
		ARLM    & Autoregressive linear mixed                                   & -       & 5.1.2  & effect of covariates          & unit\\ 
		\hline
		ARP     & Autoregressive panel data                                     & -       & \ref{sec:ARP}   & intercept                     & unit\\
		\hline
		BVAR    & Bayesian vector autoregressive                                & -       & 5.2   & AR coef                       & unit\\
		\hline
		CHARMA  & Conditional heteroscedasticity autoregressive moving average  & -       & 4.4 & AR coef                       & time \\ 
		\hline 
		DFM     & Dynamic factor                                                & -       & 6.2  & AR coef                        & time \\ 
		\hline 
		GARCH   & Generalized autoregressive conditional heteroscedasticity     & -       & 4.3 & -                             & time \\ 
		\hline
		GMB     & Generalized Markovian bilinear                                & -       & 4.1.2  & -                             & time \\ 
		\hline
		GRCA    & Generalized random coefficient autoregressive                 & -       & 4.1   & AR coef                       & time \\ 
		\hline
		HVAR    & Hierarchical vector autoregressive                            & new     & 5.3     & AR coef, effect of covariates & unit\\
		\hline
		RAR     & Random autoregressive                                         & new     &3     & AR coef, effect of covariates & time and unit\\ 
		\hline
		RARMA   & Random autoregressive moving average                          & new     &3   & AR coef, effect of covariates & time and unit\\ 
		\hline
		RCA     & Random coefficient autoregressive                             & -       &4.1.1    & AR coef                       & time \\ 
		\hline
		RCAC    & Random coefficient autoregressive with correlated terms       & new     & 4.2  & AR coef & time\\ 
		\hline
		RCAP    & Random coefficient autoregressive panel data                  & add \itshape panel data \upshape to distinguish from RCA & 5.1.1 & AR coef & unit \\ 
		\hline
		RCARRS  & Random coefficient autoregressive regime switching            & -       & 4.5 & AR coef & time \\ 
		\hline
		RCEA    & Random coefficient exponential autoregressive                 & -       & 4.1.3   & -       & time \\ 
		\hline
		RCP    & Random coefficient panel                                       & -       & 6.1  & AR coef & time and unit\\
		\hline
		TSCS   & Time-series-cross-sectional                                    & -       &5.1.3   & AR coef & unit\\ 
		\hline
		TVAR    & Time-varying autoregressive                                   & -       &4.6   & AR coef & time \\ 
		\hline
		UAR     & Unit-specific autoregressive                                  & new     & 5.1        &AR coef,  effect of covariates & unit \\ 
		\hline 
	\end{tabular}
	\caption{For each model: acronyms and corresponding full names, together with the changes made to the original name and the section in which the model is treated. Random term specifies which coefficients are random and heterogeneity specifies whether the coefficient is random in time and/or across units.}
	\label{Table:Acronyms}
\end{sidewaystable}

\newpage
\definecolor{mycolor1}{rgb}{0,0,0.7}   
\definecolor{mycolor2}{rgb}{0.7,0,0}   

{
	\centering
	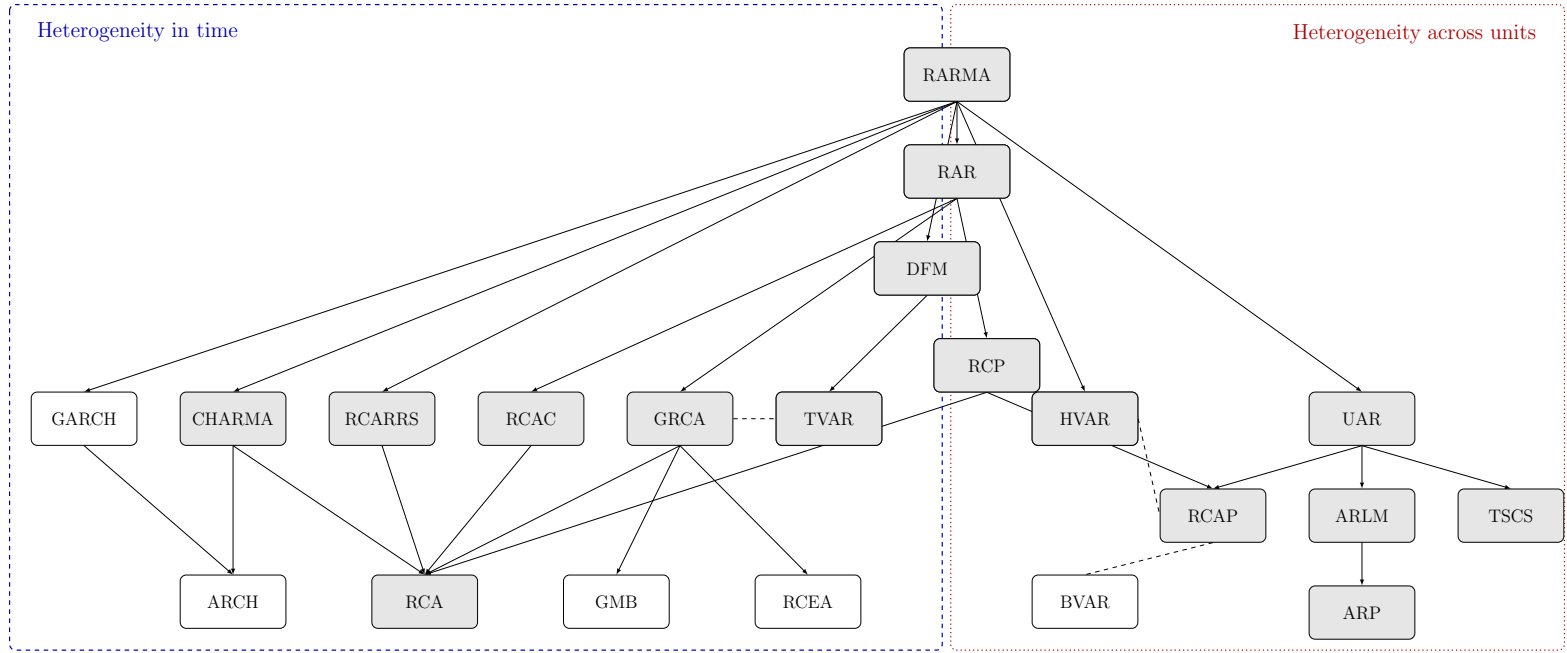
\begin{sidewaysfigure}
		\resizebox {\textwidth} {!}{
			\begin{tikzpicture}[C0/.style={rectangle, minimum height=7em, minimum width=19em}, C1/.style={rectangle, minimum height=3em, minimum width=14em}, C2/.style={rectangle, minimum height=3em, minimum width=3em}, C3/.style={rectangle, minimum height=3em, minimum width=.1pt},block/.style={draw, fill=gray!20, rectangle, rounded corners, minimum height=3em, minimum width=6em},
			block2/.style={draw, fill=white, rectangle, rounded corners, minimum height=3em, minimum width=6em},
			blockGray/.style={fill=gray!10, text=gray, rectangle, rounded corners, minimum height=3em, minimum width=6em},
			block2Gray/.style={draw, gray, fill=white, text=gray, rectangle, rounded corners, minimum height=3em, minimum width=6em}]
			
			\node[block]at (0,0) (RARMA){RARMA};
			\node[block, below = of RARMA] (RAR) {RAR};
			\node[block, below = of RAR, xshift=-0.7cm] (DFM){DFM};
			\node[block, below = of DFM, xshift=1.4cm](RCP){RCP};
			
			\node[block]at(-3cm,-8cm)(TVAR){TVAR};
			\node[block, left = of TVAR](GRCA){GRCA};
			\node[block, left = of GRCA](RCAC){RCAC};
			\node[block, left = of RCAC](RCARRS){RCARRS};
			\node[block, left = of RCARRS](CHARMA){CHARMA};
			\node[block2, left = of CHARMA](GARCH){GARCH};
			\node[block2, below = of CHARMA, yshift=-2cm](ARCH){ARCH};
			\node[block, right = of ARCH, xshift=1cm](RCA){RCA};
			\node[block2, right = of RCA, xshift=1cm](GMB){GMB};
			\node[block2, right = of GMB, xshift=1cm](RCEA){RCEA};
			
			\node[block]at(3cm,-8cm)(HVAR){HVAR};
			\node[block, right=of HVAR, xshift=3cm](UAR){UAR};
			\node[block, below =of UAR](ARLM){ARLM};
			\node[block, below =of ARLM](ARP){ARP};
			\node[block, right =of ARLM](TSCS){TSCS};
			\node[block, left  =of ARLM](RCAP){RCAP};
			\node[block2, below = of HVAR, yshift=-2cm](BVAR){BVAR};
			
			\draw[-latex](RARMA.south)--(DFM.north);
			\draw[-latex](RARMA.south)--(RAR.north);
			\draw[-latex](RARMA.south)--(GARCH.north);
			\draw[-latex](RARMA.south)--(CHARMA.north);
			\draw[-latex](RARMA.south)--(RCARRS.north);
			\draw[-latex](RAR.south)--(RCAC.north);
			\draw[-latex](RAR.south)--(GRCA.north);
			\draw[-latex](RAR.south)--(RCP.north);
			\draw[-latex](DFM.south)--(TVAR.north);
			\draw[-latex](GARCH.south)--(ARCH.north);
			\draw[-latex](CHARMA.south)--(ARCH.north);
			\draw[-latex](CHARMA.south)--(RCA.north);
			\draw[-latex](RCARRS.south)--(RCA.north);
			\draw[-latex](RCAC.south)--(RCA.north);
			\draw[-latex](GRCA.south)--(RCA.north);
			\draw[-latex](GRCA.south)--(RCEA.north);
			\draw[-latex](GRCA.south)--(GMB.north);
			\draw[-latex](RCP.south)--(RCA.north);
			
			\draw[-latex](RCP.south)--(RCAP.north);
			\draw[-latex](RARMA.south)--(UAR.north);
			\draw[-latex](RARMA.south)--(HVAR.north);
			
			\draw[-latex](UAR.south)--(RCAP.north);
			\draw[-latex](UAR.south)--(ARLM.north);
			\draw[-latex](UAR.south)--(TSCS.north);
			\draw[-latex](ARLM.south)--(ARP.north);
			
			\draw[dashed](GRCA.east)--(TVAR.west);
			\draw[dashed](RCAP.south)--(BVAR.north);
			\draw[dashed](HVAR.east)--(RCAP.west);
			
			\draw[mycolor1, thick, dashed, rounded corners] ($(GARCH.north west)+(-0.5, +9)$)  rectangle ($(RCEA.south east)+(1.9, -0.5)$);
			\draw[mycolor2, thick, dotted, rounded corners] ($(UAR.north east)+(+3.5, +9)$)  rectangle ($(BVAR.south west)+(-1.9, -0.5)$);
			
			\node[block]at (0,0) (RARMA){RARMA};
			\node[block, below = of RARMA] (RAR) {RAR};
			\node[block, below = of RAR, xshift=-0.7cm] (DFM){DFM};
			\node[block, below = of DFM, xshift=1.4cm](RCP){RCP};
			\node[block]at(3cm,-8cm)(HVAR){HVAR};
			\node[block]at(-3cm,-8cm)(TVAR){TVAR};

			\node[above right, text=mycolor1] at ($(GARCH.north west)+(0, +8)$) {\large Heterogeneity in time};
			\node[above right, text=mycolor2] at ($(TSCS.north east)+(-6.5,+10.2)$) {\large Heterogeneity across units};
			\end{tikzpicture}
			
		}
		\caption{{Hierarchical structure of model families grouped by the type of heterogeneity they are suited for.}\newline The blue dashed line groups models for heterogeneity in time. The red dotted line clusters models for heterogeneity across units.
			The acronyms in the graph's node correspond to the full model specifications provided in Table~\ref{Table:Acronyms}. The arrows point from the general structure to the particular case, while the dashed lines connect similar models in a broader sense. The white nodes are shortly treated in this overview, as they do not fulfil the inclusion criteria of the present work.}\label{fig:structure}
	\end{sidewaysfigure}
}

\newpage
\section{Random autoregressive models}\label{sec:RAR}
An $m$-variate time series $\{\YY{i}{t}\}$ for unit $i=1,\dots, n$ at time $t=1,\dots,T_i$ is said to follow a random autoregressive model of order $p$ -- RAR($p$) if it satisfies
\begin{equation}
\YY{i}{t} = \sum_{k=1}^{p} \rYtilde{i}{t}{k} \YY{i}{t-k} + \design{X}{i}{t} \beta + \design{Z}{i}{t}\rZ{i}{} + \ee{i}{t},\label{eq:RAR}
\end{equation}
where $\YY{i}{t}$ and $\YY{i}{t-k}$ are the $m$-vectors of response variables for unit $i$ at time $t$ and $t-k$ respectively.
The $m\times mr$ and $m\times ms$ block diagonal matrices $\design{X}{i}{t}$ and $\design{Z}{i}{t}$ are the design matrices  at time $t$ for the fixed and random effects respectively.
The $mr$-vector $\fX = [\fX_1^T, \dots, \fX_m^T]^T$ and the $ms$-vector $\rZ{i}{} = [\rZ{i}{1}^T, \dots, \rZ{i}{m}^T]^T$ represent fixed and random effects, with $\fX_j$ and $\rZ{i}{j}$ the vectors corresponding to the $j$th outcome. In particular, $\rZ{i}{}$ is sampled from a distribution with zero mean and constant covariance matrix $\Sigma_c$.
The $m\times mp$-matrices of auto- and cross-regression coefficients $\{ \rYtilde{i}{t}{}\}$ constitute a stochastic process indexed by $i$ and $t$, where $\rYtilde{i}{t}{}=[\rYtilde{i}{t}{1},\dots,\rYtilde{i}{t}{p}]$.
Finally, the noise process $\{\ee{i}{t}\}$ is a sequence of $m$-variate random vectors with mean zero and covariance matrix $\Sigma_{e_{i,t}}$.
No assumption is made on serial and mutual correlation between $\{\rYtilde{i}{t}{}\}$ and $\{\ee{i}{t}\}$, while all the random variables are independent across units.
The random effects $\rZ{i}{}$ are also independent of all other random coefficients,
\begin{equation*}
\text{Cov}
\begin{pmatrix}
\rYtilde{i}{t}{} \\
\rZ{i}{}\\
\ee{i}{t}
\end{pmatrix}=
\begin{pmatrix}
\Sigma_{g_i}  & 0           & \Sigma_{ge}\\
0             & \Sigma_c    & 0\\
\Sigma_{ge}   & 0           & \Sigma_{e_{i,t}}
\end{pmatrix},
\end{equation*}
with $\Sigma_{ge}$ a covariance matrix. Furthermore, the initial state $\YY{i}{0 }$ is assumed to be of finite variance for any $i$.

We make here a brief comment about the existence and identifiability of processes such as~\eqref{eq:RAR}.
Following~\cite{Nicholls1982} and~\cite{Lutkepohl2005}, one can consider the simpler case of a first-order process on one unit since a vector autoregressive (VAR) process of general order $p$ can always be written as a VAR(1) model~\citep{Lutkepohl2005}:
\[
\YY{t}{} = \rYtilde{t}{}{} \YY{t-1}{}  + \design{X}{t}{} \fX + \design{Z}{t}{} \rZ{i}{} + \ee{t}{}.
\]
Without loss of generality we can set $\beta=0$, and the random effects can be studied separately.
It remains therefore to examine the process
\[
\YY{t}{} = \rYtilde{t}{}{} \YY{t-1}{}  + e_t.
\]
This is a VAR(1) model with a random autoregressive coefficient, so this representation ensures that if the coefficients $\rYtilde{t}{}{}$ almost surely satisfy certain conditions (namely those under which VAR processes exist), then also~\eqref{eq:RAR} will exist.
Note however that~\eqref{eq:RAR} defines a much wider class of processes.
Attesting to results in the literature, the existence of particular instances of~\eqref{eq:RAR} has already been studied:
\cite{Vanvevcek2007} considers the case in which the autoregressive coefficients are serially uncorrelated, and uncorrelated with the residuals;
\cite{Lutkepohl2005} assumes the autoregressive coefficients to satisfy almost surely the conditions for existence of VAR models, and to be uncorrelated both serially and of the residuals.

Model~\eqref{eq:RAR} in its full specification is probably overparametrized.
We obtain meaningful submodels by introducing constraints on the model parameters.
For instance, assume that there exists a filtration $\mathcal{F}_t$ such that
\begin{align}
\E {\rYtilde{i}{t}{} |\mathcal{F}_t} &= \rY{i}{}{},\text{or} \label{eq:a}\\
\E {\rYtilde{i}{t}{} |\mathcal{F}_t} &= \rY{t}{}{}, \text{or}\label{eq:b}\\
\E {\rYtilde{i}{t}{} |\mathcal{F}_t} &= \fY{}{}.\label{eq:c}
\end{align}
In these cases, the resulting autoregressive coefficient, although being random, presents only subject-specific variability~\eqref{eq:a}, or the way it changes in time is the same across subjects~\eqref{eq:b}, or it does not depend on either unit $i$ or time $t$~\eqref{eq:c}.
Many other assumptions can be stated to derive existence and identifiability of the process~\eqref{eq:RAR}, but this is outside the scope of the present work.

In order to have a structure that includes all the models treated in the present work as particular cases, we extend the definition of RAR models to random autoregressive moving average models -- RARMA($p,q,r,s$) -- with a state-space formulation very close to the one of~\cite{Tsay1987} (see Figure~\ref{fig:structure}) 
\begin{equation}
\begin{aligned}
\YY{i}{t} &= \sum_{k=1}^{p}\fY{k}{} \YY{i}{t-k} + \sum_{k=1}^{q} \fe{k}{} \ww{i}{t-k} + \design{X}{i}{t} \fX + \design{Z}{i}{t} \rZ{i}{t} + \ww{i}{t}\\
\ww{i}{t} &= \sum_{j=1}^{r} \re{i}{t}{j} \ww{i}{t-j}  +\sum_{j=1}^{s} \rY{i}{t}{j} \YY{i}{t-j} +\rforecast{i}{t} \YYhat{i}{t}+\ee{i}{t}. \label{eq:RARMA}
\end{aligned}
\end{equation}
Here $\fY{}{}=[\fY{1}{},\dots,\fY{p}{}]^T$ and $\fe{}{}=[\fe{1}{},\dots,\fe{q}{}]^T$ are constant autoregressive and moving average coefficients respectively, $\re{i}{t}{}=[\re{i}{t}{1},\dots,\re{i}{t}{r}]^T$, $\rY{i}{t}{}=[\rY{i}{t}{1},\dots,\rY{i}{t}{s}]^T$ and $\rforecast{i}{t}$ are stochastic processes indexed by $i$ and $t$ with mean zero, and 
\begin{equation*}
\text{Cov}
\begin{pmatrix}
\rZ{i}{t}\\
\rY{i}{t}{} \\
\re{i}{t}{}\\
\rforecast{i}{t}\\
\ee{i}{t}
\end{pmatrix}=
\begin{pmatrix}
\Sigma_c  & 0         & 0         &  0        & 0 \\
0         & \Sigma_{g_i}  & 0         &  0        & \Sigma_{ge}\\
0         & 0         & \Sigma_f  &  0        & 0\\
0         & 0         & 0         &  \Sigma_l & 0\\
0         &\Sigma_{ge}& 0         &  0        & \Sigma_{e_{i,t}}\\
\end{pmatrix}.
\end{equation*}
Furthermore, $ \YYhat{i}{t-1}$ denotes the conditional expectation of $\YY{i}{t}$ given $\mathcal{F}_{t-1}$.
Stability and stationarity of the process in~\eqref{eq:RARMA} have been proven by~\cite{Tsay1987} when some filtration satisfying~\eqref{eq:b} is imposed, and $\design{X}{i}{t}=\design{Z}{i}{t}=0\ \forall i,t$. Note that a RARMA($p,0,0,p$) model with $\rforecast{i}{t}{}=0$ a.s.~is in fact a RAR($p$) model, where $\rYtilde{i}{t}{k} = \rY{i}{t}{k} + \fY{i}{k}$.
Similarly we will show that also the other models presented in this overview are particular cases of the general RARMA structure (see Figure~\ref{fig:structure}).

\section{Models for the heterogeneity in time}\label{sec:TimeHeterogen}
This section deals with models suited for $(1,m,T)-$data, typically with $T$ large, grouped by the blue dashed box on the left hand side of Figure~\ref{fig:structure},
and characterized by parameters that are random in time.
Since in all models treated here $n=1$, we drop the subscript $i$.
It is worth mentioning that such models have also been treated in the  closely related field of statistical tracking, but we do not pursue this connection here (c.f.~\cite{Kushner2003} for an overview).

\subsection{Generalized random coefficient autoregressive models (GRCA)}\label{sec:GRCA}
With the notation and distributional assumptions introduced in Section~\ref{sec:RAR}, the generalized random coefficient autoregressive (GRCA) model of order $p$ can be written as
\begin{equation}
\YY{t}{}=\sum_{k=1}^{p}\left( \fY{k}{}+\rY{t}{k}{} \right) \YY{t-k}{} +\ee{t}{}. \label{eq:GRCA}
\end{equation}
It is in fact a RARMA($p,0,0,p$) model with $\rforecast{t}{}=0$ a.s. and $\design{X}{t}{}=\design{Z}{t}{}=0$ for every $t$. The autoregressive coefficients $\rY{t}{}{}=[\rY{t}{1}{},\dots,\rY{t}{p}{}]^T$ and errors $\ee{t}{}$ are serially uncorrelated, but they are allowed to be mutually correlated.
In the literature GRCA models are often referred to as random coefficient autoregressive (RCA) models~\citep{Conlisk1974, Conlisk1976,Chandra2001,Hill2014}, although RCA models are in fact a particular case of GRCAs (c.f.~Section~\ref{sec:RCA})).
The GRCA includes also the (generalized) Markovian bilinear ((G)MB) model and the random coefficient exponential autoregressive (RCEA) models for particular choices of the autoregressive coefficients~\citep{Hwang1998}.
Such models have been developed for econometric and financial {\bf applications}, but have been later inherited in engineering (e.g.~hydrology, metrology and telecommunication) and biology (e.g.~dynamic population models) for their flexibility in modelling occasional sharp spikes. 

Necessary and sufficient conditions for {\bf stability}  of GRCA models were derived in early references by~\cite{Conlisk1974, Conlisk1976}.
For {\bf inference} on $\gamma$ various methods have been proposed: maximum likelihood, shown to lead to locally asymptotically normal and asymptotically optimal estimators~\citep{Hwang1997}, conditional and weighted conditional least squares (CLS and WLS)~\citep{Hwang1998}, and
the estimated optimal estimators (EOE) of~\citeauthor{Chandra2001} combining Godambe's optimal estimating functions with CLS and the method of moments.
In terms of efficiency, WLS has been shown to outperform CLS~\citep{Hwang1998}, and to be identical to the EOE~\citep{Chandra2001}. 

More recent publications have proposed estimation methods that can estimate both $\gamma$ and the covariance parameters, previously treated as nuisance.
\cite{Fink2014} propose bootstrap methods based on quasi-maximum likelihood (QML), while~\cite{Zhao2012} introduce empirical likelihood (EL) estimators, deriving also the asymptotic distribution of the estimators and a non-parametric version of Wilk's theorem.
The advantages of EL on WLS include robustness against heteroskedasticity, distribution-adapted confidence intervals and avoiding the estimation of the asymptotic covariance matrix.

For the {\bf estimation of non-stationary GRCAs} (and analogously for generalized autoregressive conditional heteroskedasticity (GARCH) models, c.f.~Section~\ref{sec:GARCH}), \cite{Truquet2012} introduce QML estimation, and prove its asymptotic normality and consistency.
\cite{Hill2014} create a unified framework in the estimation of GRCA for stationary and non-stationary, possibly trended time series, either with or without random coefficients. This is achieved by combining EL with the WLS score equation. 

The literature on {\bf tests for GRCA}, moreover, includes tests for stationarity and ergodicity in GRCA(1)~\citep{Zhao2012, Zhao2015}, 
and for parameters change~\citep{Zhao2013, Zhao2014}.
In another publication an approach to variable selection is also proposed~\citep{Zhao2018}.

\subsubsection{Random coefficient autoregressive models (RCA)}\label{sec:RCA}
Assuming $\rY{t}{}{}$ to be independent of the error $\ee{t}{}$ in~\eqref{eq:GRCA}, this equation leads to the random coefficient autoregressive (RCA) model of~\cite{Nicholls1982}, as previously pointed out by~\cite{Hwang1998}.
This structure constitutes one of the first attempts to capture the heterogeneity typical of time series in financial {\bf applications}. As for GRCA, this model has then been used in many other fields where random perturbations are present, such as in macroeconomy, engineering (e.g.~car vibrations or ship rolling), biology (e.g.~brain-waves recordings) and meteorology.

{\bf Stability, stationarity and ergodicity} in RCA models have been largely studied in early publications by~\cite{Andvel1976, Nicholls1981, Nicholls1982, Feigin1985}.
Stability and ergodicity are then classical assumptions when dealing with {\bf estimation of RCA}.
Least squares estimation has been largely used~\citep{Nicholls1981, Nicholls1982, Tsay1987}, and is often taken as benchmark for modern advances.
Furthermore, since LS estimates are strongly consistent and under suitable conditions they obey the central limit theorem, they often serve as starting point for iterative schemes such as maximum likelihood (ML)~\citep{Nicholls1982, Tjostheim1986, Allal2013}, or in combination with other procedures such as estimating functions (EF)~\citep{Thavaneswaran1988, Abdullah2011} and bootstrap methods~\citep{Pravskova2003, Fink2013}.
Note that for ML estimation, autoregressive coefficients and residuals are also typically assumed to be jointly normal.
Furthermore, for RCA models, the EF is equivalent to WLS~\citep{Abdullah2011}.
\cite{Schick1996} and~\cite{Koul1996} 
derive an efficient adaptive locally asymptotically minimax estimator for $\gamma$.
\cite{Qian1996} propose minimum distance estimators, while~\cite{Ghahramani2009} combine LS's and least absolute deviations (LAD)'s estimating functions to estimate model volatility (both for RCA and for GARCH models, c.f.~\ref{sec:GARCH}).
\cite{Aue2006} use QML to estimate the parameters of a scalar RCA(1), and impose only minimal conditions on the sequences of random coefficients and residuals. Under suitable conditions, they derive strong consistency and asymptotic normality for these estimates.
Recently the generalized moment estimator and the Whittle estimator (involving an approximation to the likelihood function) have also been proposed~\citep{Shitan2015, Bibi2016}.

{Inference} for models that fall under the RARMA structure normally requires favourable assumptions, such as stationarity and limited second order moments. 
However, especially for RCA models, the literature is broad and deals also with non-standard assumptions. 
Authors have explored {\bf estimation of non-stationary RCAs}, and as for the GRCA model, proposed QML estimation.
Although the variance of the initial error cannot be estimated, weak consistency and asymptotic normality of QML estimates can be proven~\citep{Berkes2009, Aue2011}.
In the latter reference, the method is validated through a Monte Carlo simulation study and applied to real financial data.
The Bayesian approach also enables the analysis of non-stationary time series, and allows the inclusion of prior information~\citep{Diaz1990, Yang1995, Barnett1996, Safadi2003, Wang2008, Wang2009, De2014}.
The comparison between the Bayesian approach of~\cite{Wang2009} and the QML approach applied to the same daily stock transaction volume data leads to a non-decisive result on the real data-set, but to the conclusion of a better performance of the frequentist approach on the simulated data~\citep{Aue2011}.

For RCAs, authors have also considered the case of {\bf  infinite error variance}, and proposed (smoothed) LAD for inference, consistent and asymptotically normal~\citep{Thavaneswaran2001, Thavaneswaran2004}, and providing more precise estimates than least squares in the case of infinite variance~\citep{Goryainov2016}.
The CLS estimator of the autoregressive coefficient is also shown to be asymptotically normal when the second moment of the innovation is infinite under some weak conditions~\citep{Fu2015}.

Recently a {\bf test} for strict stationarity of RCAs has been proposed by~\cite{Trapani2020}, while other tests exist for parameter changes~\citep{Pravskova2015, Li2015, Li2015a}, and for the randomness of the coefficients~\citep{Lee1998, Akharif2003, Horvath2019} -- the latter publication in particular tests for a null on the boundary of the parameter space.

In econometrics, {\bf testing for unit root} in autoregressive models is of particular interest~\citep{Nelson1982, Phillips1988}.
This naturally extends to RCA models, where the unit root assumption reduces to $\gamma=1$, together with the alternatives of near-stationarity ($\gamma \rightarrow 1 $) and mild explosivity ($\gamma > 1$)~\citep{Leybourne1996, Sollis2000,Aue2008}, and it closely relates to the detection of bubbles~\citep{Psaradakis2001, Banerjee2020}.
Many references report tests, inference methods and asymptotic consequences of the three regimes, determined by how $\gamma$ approaches the unit value~\citep{Wang2008, Nagakura2009, Nagakura2009b, Aue2008}.
Note furthermore that also in this case, there exists a completely separated literature dealing with the so-called stochastic unit root (STUR) model, explosive random coefficient autoregressive (ERCA) model, or near-explosive random coefficient autoregressive model (NERC) with very different acronyms while in fact all are RCAs with particular $\gamma$'s.

\subsubsection{Generalized Markovian Bilinear models (GMB)}\label{sec:GMB}
Take $\rY{t}{k}{} = \theta_k \ee{t}{}^{r_k}$ in~\eqref{eq:GRCA}, with $\theta=[\theta_1, \dots, \theta_p]^T$ a vector of constants and $r=[r_1,\dots,r_p]^T$ a vector of non-negative integers.
The resulting model is the so-called generalized Markovian bilinear (GMB) model,
\begin{equation}
\YY{t}{}=\sum_{k=1}^{p} \left(\fY{k}{} + \theta_k \ee{t}{}^{r_k}\right) \YY{t-k}{} + \ee{t}{} \label{eq:GMB},
\end{equation}
of which the Markovian bilinear model is a particular case when $r_k=1\ \forall k$~\citep{Tong1981,Feigin1985,Cline2002}.
Note that although GMB models have a random autoregressive coefficient, this coefficient is not sampled from a different stochastic process, as it is for RCA models (c.f.~Section~\ref{sec:RCA}).
Rather, the autoregressive coefficient is function of the noise term.

\subsubsection{Random Coefficient Exponential Autoregressive models (RCEA)}\label{sec:RCEA}
Take $\rY{t}{k}{} =  (\theta_{k_1} + \theta_{k_2} \exp(-\theta_{k_3}e_t^2) )e_t$ in~\eqref{eq:GRCA}, where $\theta_j=[\theta_{1_j}, \dots, \theta_{p_j}]^T,\ j=1,2,3$ are vectors of constant parameters. Then, \eqref{eq:GRCA} becomes
\begin{equation}
Y_{t}=\sum_{k=1}^{p} \left( \fY{k}{} + (\theta_{k_1} + \theta_{k_2} \exp(-\theta_{k_3}\ee{t}{}^2) )e_t\right) \YY{t-k}{} + \ee{t}{} \label{eq:RCEA},
\end{equation}
known in literature as random coefficient exponential autoregressive (RCEA) model~\citep{Hwang1998, Priestley1980}.
As in GMB models, the heteroskedasticity in the autoregressive coefficient is inherited from the error term.

\subsection{Random coefficient autoregressive models with correlated terms (RCAC)}\label{sec:RCAC}
In practical {\bf applications} for example encountered in economy and finance (e.g.~volume transaction data), the assumption of independence in time of the coefficients often does not hold.
Models that account for some time-dependence among the random parameters have thus been introduced.
These models also extend the RCA model (Section~\ref{sec:RCA}), but are not particular cases of the GRCA. In fact they are again derived from~\eqref{eq:RAR} without the subscript for the unit, by setting $\design{X}{t}{}= \design{Z}{t}{}=0\ \forall t$, but contrarily to GRCA they assume independence between noise and autoregressive coefficients, and allow serial correlation for one of the two random variables.
The acronym introduced here for this family is RCAC -- random coefficient autoregressive models with correlated terms.

\subsubsection{Random coefficient autoregressive model with correlated random coefficients}
An example of RCAC model relaxes the assumption of serial independence of the autoregressive coefficients, by assuming the random coefficient of the scalar RCA(1) to be $\rY{t}{1}{}= \alpha_0 z_t + \alpha_1 z_{t-1}$,
with $\alpha_0, \alpha_1\neq 0$ constant coefficients and $z_1,\dots, z_t$ independent random variables with zero means and the same variance $\sigma^2_z$, independent of both $\YY{0}{}$ and $\{\ee{t}{}\}$~\citep{Koubkova1982}.
In this publication, conditions for {\bf stationarity}, covariance function and best linear predictions are obtained.

\subsubsection{Random coefficient autoregressive model with correlated error sequence}
In the dissertation, \cite{Vanvevcek2007} extends the results for RCA models under typical assumptions (c.f.~Section~\ref{sec:RCA}) to RCAs with the noise being an ergodic, strictly stationary martingale difference sequence  with respect to the previous observations. 
Although the author refers to it as a generalized random coefficient autoregressive model, this should not be confused with model~\eqref{eq:GRCA} as pointed out in the second chapter of the dissertation, since the autoregressive coefficient is independent of the residuals.
The author proposes a new functional {\bf estimator} for RCA$(p)$ and multivariate RCA$(1)$ under this assumption, extending the method originally proposed by~\cite{Schick1996} for RCA$(1)$ models.
The newly introduced estimator is strongly consistent and asymptotically normal.
With an extensive simulation study the author compares LS, WLS, ML and the new functional estimator in terms of efficiency and asymptotic variance, and concludes that the WLS seems the optimal choice.
The code for parts of the {\bf implementation in R}~\citep{Rmanual} is provided in the publication.
A study comparing the performance of the LS method on RCA models with correlated and uncorrelated error sequence is presented by~\cite{Araveeporn2013}. From the simulation study, the author concludes that accounting for correlation in residuals improves the results only when the data oscillates, and for real data the model with autocorrelations outperforms the simpler model with serially independent error sequence.

\subsection{Generalized autoregressive conditional heteroskedasticity models (GARCH)}\label{sec:GARCH}
A significant contribution to the literature for modeling stochastic volatility in financial {\bf applications} is given by the family of generalized autoregressive conditional heteroskedasticity models of~\cite{Bollerslev1986}.
With the notation of the previous sections a GARCH$(p,q)$ model is defined as
\begin{equation}
\begin{aligned}
\YY{t}{}|\mathcal{F}_{t-1} &\sim N(X\beta, h^2_t)\\
h^2_t &= \alpha_0 + \sum_{k=1}^{p}\alpha_k\ww{t-k}{}^2 + \sum_{k=1}^{q}\beta_k h^2_{t-k}\\
\ww{t}{} &= \YY{t}{} - X\beta
\label{eq:GARCH}
\end{aligned}
\end{equation}
where $\mathcal{F}_{t-1}$ is a filtration, and $\alpha_0$, $\alpha_k$, and $\beta_k$ are coefficients to be estimated.
This model is suited for~$(1,m,T)-$data, with $m$ possibly larger than one~\citep{Bauwens2006}.
This model can also be seen as a RARMA($0,0,0,0$) model with $\design{Z}{t}{}= 0\ \forall t$ and $\rforecast{t}{}=0$ almost surely. Thus $\ww{t}{} = \ee{t}{}$, and $\ee{t}{}$ is normally distributed with variance $h^2_t$.
As for GMB and RCEA models (Sections~\ref{sec:GMB} and~\ref{sec:RCEA}), the heteroskedasticity in time comes from the noise and not from sampling from a different distribution. We mention this family shortly for completeness, and refer to other references that deal in depth with this model~\citep{Bauwens2006, Francq2011}, and mention~\cite{Mccullough1998, Brooks1997, Brooks2003, Ghalanos2014, Ghalanos2019, GARCH_Matlab, Kim1993, Boffelli2016} for what concerns {\bf software implementation}.

\paragraph{Autoregressive conditional heteroskedasticity models (ARCH)}\label{sec:ARCH}
The particular case of~\eqref{eq:GARCH} with $q=0$ constitutes the autoregressive conditional heteroskedasticity (ARCH) model of order $p$ of~\cite{Engle1982}.
Here the covariance structure is only function of the noise at previous states,
$
h^2_t = \alpha_0 + \sum_{k=1}^{p}\alpha_k\ww{t-k}{}^2 ,
$
imposing shorter memory to the process.

Similarly to GARCH the heterogeneity in ARCH is obtained by incorporating the error term in the definition of some parameters, and we refer to publications specifically on the topic for further details~\citep{Bollerslev1992, Degiannakis2004,Gourieroux2012}, while for {\bf software packages} one can refer to the ones mentioned for fitting GARCH models (c.f.~Section~\ref{sec:GARCH}).
It is however interesting to note that RCA and ARCH models share second order properties and can be studied in parallel~\citep{Tsay1987, Wolff1988,Chandra2001}.

\subsection{Conditional heteroskedasticity autoregressive moving average models (CHARMA)}\label{sec:CHARMA}
After the introduction of the ARCH model (Section~\ref{sec:ARCH}), many authors investigated its relationship with the RCA model (Section~\ref{sec:RCA}), and their contribution in econometric {\bf applications}, such as in spot rate predictions.
In fact, the first can be rewritten with the structure of the second, and they possess the same second order properties~\citep{Tsay1987, Wolff1988}.
The introduction of conditional heteroskedasticity autoregressive moving average (CHARMA) models by~\cite{Tsay1987} is motivated by the consideration that neither RCA nor ARCH have parsimonious definitions.
CHARMA models are defined by the equations
\begin{equation}
\begin{aligned}
\fY{}{}(B)\YY{t}{}    &= \fe{}{}(B)\ww{t}{} \\
\re{t}{}{}(B)\ww{t}{} &= \rforecast{t}{} \YYhat{t-1}{} + \rY{t}{}{}(B) Y_t + \ee{t}{}
\end{aligned}
\end{equation}
where $B$ is the backwards operator, and are in fact RARMA($p,q,r,s$) with $\design{X}{t}{} = \design{Z}{t}{} =0,\ \forall t$, and with both serially and mutually independent random terms having constant variances.
{\bf Estimation} can be carried out via LS, and appropriate moment conditions ensure asymptotic normality of the estimators.
Multivariate extensions of the CHARMA model are discussed by~\cite{ahn1990estimation}.
Note that RCA and ARCH are also particular cases of CHARMA~\citep{Tsay1987}.

\subsection{Random coefficient autoregressive regime switching models (RCARRS)}\label{sec:RCARRS}
For the particular {\bf application} of estimating time-varying hedges ratios, many models have been explored --
in particular, under the RARMA family, the bivariate GARCH and the RCA models~\citep{Bera1997}.
The so-called random coefficient autoregressive regime switching (RCARRS) model of~\cite{Lee2006} was introduced by combining the flexibility of RCA models with the state-dependence properties of Markov regime switching (MRS) models~\citep{Alizadeh2004}. It was originally formulated as
\begin{align}\label{eq:RCARRS}
S_t &= \alpha_{s_t} + \zeta_t F_t + \varepsilon_{t,s_t} \\
(\zeta_{t}-\bar{\zeta})   &= \phi (\zeta_{t-1}-\bar{\zeta})  + \nu_t,
\end{align}
where $S_t$ and $F_t$ are the spot and future values at time $t$, $\alpha_{s_t}$ is a state-dependent constant parameter, $\zeta_{t}$ is the autoregressive coefficient with steady-state $\bar{\zeta}$. The two iid sequences $\varepsilon_{t,s_t}$ and $\nu_t$ with variances $\sigma_{\varepsilon_{s_t}}^2$ and $\sigma^2_{\nu}$ respectively are also mutually independent.
The state equation on $\zeta_{t}$ has autoregressive coefficient $|\phi|<1$ to impose stationarity on the $\zeta_{t}$ sequence.
The parameters depend on a latent variable $s_t$ following a two-state first-order Markov-switching process with unknown transition probabilities $p_1$ and $p_2$. For {\bf estimation}, \citeauthor{Lee2006} propose ML while recurring to Kim’s filter (\cite{Kim1994}; a combination of the extended Kalman and Hamilton filters) after an appropriate transformation. 
From a comparison of the models proposed for this application, \cite{Lee2006} conclude that RCA has the best in-sample (on the data used to estimate model parameters) performance, while RCARRS the best out-of-sample (on new data) performance for the case study at hand.
With a small extension to state-dependent variables, the RCARRS model of order $p$ can be seen as a RARMA($1,1,0,0$) model with $\fY{k}{}=\phi_k$, $\fe{k}{}=-\phi_k$, $\design{X}{t}{} = [1, F_t]$, $\design{Z}{t}{} =  F_t$, $\fX=[(1-\phi)\alpha_{s_t}, \bar{\beta}]^T$ and $\rZ{t}{}=\nu_t$.
Furthermore, $s_t \in \{0,1\}$ determines the distribution being sampled,
\begin{align*}
\alpha_{s_t} &= \alpha_{0}\mathbbm{1}\{s_t=0\} + \alpha_{1}\mathbbm{1}\{s_t=1\},\\
\varepsilon_{t,s_t} &= \varepsilon_{t,0}\mathbbm{1}\{s_t=0\} +  \varepsilon_{t,1}\mathbbm{1}\{s_t=1\},
\end{align*}
where $\alpha_{0}$ and $\alpha_{1}$ are two constant values, $\varepsilon_{t,0}\sim N(0,\sigma_0^2)$ and $\varepsilon_{t,1}\sim N(0,\sigma_1^2)$.

\subsection{Time-varying autoregressive models (TVAR)}\label{sec:TVAR}
The stream of publications on Bayesian time-varying autoregressive (TVAR) models~\citep{Prado1997, Prado2000} developed completely independently from the remaining RARMA literature. The Bayesian framework, together with the non-stationary time series in hand, may be the reason for the non-existence of references to other publications also dealing with heteroskedastic time series treated in other sections.
In fact, this dynamic linear regression model can be derived from the general equation~\eqref{eq:RAR} similarly to GRCA models. 
On the other hand, TVAR models allow order uncertainty and multiple assumptions on the random parameters.
Such models find {\bf application} in various modern contexts, such as biomedical signal processing (including the analysis of multiple electroencephalographic traces) and communications.
\cite{Prado2000} consider a model for $(1,1,T)-$data with the autoregressive coefficient following a Gaussian random walk. 
The variance of the noise term $\sigma_t^2=\sigma_{t-1}^2(\delta/\eta_t)$ depends on $\eta_t\sim Beta(a_t, b_t)$, allowing for heteroskedasticity. Furthermore, $\eta_t$ is assumed to be serially independent and independent of $e_t$ and $\zeta_t$.
Also this model is a special case of our general structure. With the mentioned choices, it is a RARMA($0,0,0,1$) model with $\design{X}{t}{}=\design{Z}{t}{}=0\ \forall t$ and $\rforecast{t}{}=0$ almost surely, $\rY{t}{}{}$ following a random walk and $\{e_t\}$ having variance $\Sigma_{e_{i,t}}=\sigma^2_t$ as defined in the paper and rewritten above.


\section{Models for the heterogeneity across units}\label{sec:UnitHeterogen}
In this section we describe the family of models delimited by the red dotted line on the right hand side of Figure~\ref{fig:structure}.
The common feature among these models is that they pool information from $n$ time series assumed to follow the same underlying process.
This is important when the number of observations $T$ in ($n,m,T$)-data is limited.
Vector autoregressive (VAR) models are often employed to consider multiple responses simultaneously, but they are characterized by a large number of parameters which makes the estimation challenging.
Bayesian vector autoregressive (BVAR) models aim to overcome this issue by pooling information from different units, but, as noted by~\cite{Nandram1997}, they require more restricting assumptions on the priors when there is a large number of these short time series.
In fact, many of the models treated in this section were introduced to extend the flexibility of BVAR models and enable a more efficient analysis of a large number of short time series by introducing random coefficients. 
The randomness can be found either in the dynamic part (i.e.~in the autoregressive coefficient) or in the static part (i.e.~random effects), depending on the purpose.

Publications about models in this family find application in the economic, sociological, biological, agricultural, international relations and industrial fields, where multiple time series with similar behavior are available.
There exist various overviews on the topic~\citep{Franses2006, Horvath2008, Hsiao2014, Krishnakumar2012}, but their scope is limited to a single field of application and related publications.
\cite{Horvath2008} summarize various approaches specific to marketing applications, giving the appropriate estimation method, drawbacks and references for each.
All of the models that they investigate are particular cases of the RARMA structure, accommodating different levels of heterogeneity, and they are compared both via a simulation study and the application to a real case study.
\cite{Franses2006} focuses on marketing applications, and beside reviewing existing models, proposes a series of possible extensions combining well-known features -- some of which already exist in different fields of application.
Also the overview of~\cite{Hsiao2014} focuses on models for economic applications, ranging from static random effects models to those including autoregressive structure.

Since the existing literature is wide and diverse, but at the same time it shows many similarities, we introduce the unit-specific autoregressive (UAR) model to organize the right hand side of Figure~\ref{fig:structure}.
We state how to set the parameters of the UAR structure to obtain the remaining models as particular cases, showing their hierarchy and listing the main related publications.

\subsection{Unit-specific autoregressive models (UAR)}\label{sec:UAR}
With the notation of Section~\ref{sec:RAR}, the general unit-specific autoregressive (UAR) model of order $p$ can be written as 
\begin{equation}
\begin{aligned}
\YY{i}{t} &= \sum_{k=1}^{p}\fY{k}{} \YY{i}{t-k} + \design{X}{i}{t} \fX + \design{Z}{i}{t} \rZ{i}{} + \ww{i}{t},\\
\ww{i}{t} &= \sum_{j=1}^{r} \re{i}{j}{} \ww{i}{t-j}  +\sum_{j=1}^{s} \rY{i}{j}{} \YY{i}{t-j} +\ee{i}{t}. \label{eq:UAR}
\end{aligned}
\end{equation}
It is in fact a RARMA($p,0,r,s$) model where the random coefficients are indexed only by $i$ and are independent, and $\rforecast{i}{t}=0$ almost surely.

\subsubsection{Random coefficient autoregressive panel data models (RCAP)}\label{sec:RCAP}
In~\eqref{eq:UAR} take $\design{X}{i}{t}=\design{Z}{i}{t}=0\ \forall i,t$, and $\re{i}{j}{}=0$ almost surely
\begin{equation}
\begin{aligned}
\YY{i}{t} &= \sum_{k=1}^{p}\fY{k}{} \YY{i}{t-k} + \ww{i}{t},\\
\ww{i}{t} &= \sum_{j=1}^{s} \rY{i}{j}{} \YY{i}{t-j} +\ee{i}{t}. \label{eq:RCAP}
\end{aligned}
\end{equation}
The resulting equation is the random coefficient autoregressive model, which has been widely studied and applied mainly in economic and biological {\bf applications}, for its ability to pool multiple time series~\citep{Liu1980, Tiao1993}.
Since there is no agreement in the definitions, we refer to this structure as random coefficient autoregressive panel data (RCAP) model, to keep the analogy with RCA models (Section~\ref{sec:RCA}).
The residuals are assumed to be a series of identically distributed normal random variables with variance $\Sigma_{e_{i,t}}$.
In most cases the covariance matrix is diagonal and constant, but there exist exceptions~\citep{Pai1994, Nandram1997}, and only one reference makes the explicit assumption for the independence between autoregressive coefficients and error term~\citep{Safadi2003}.

{\bf Estimation for RCAP models} is often addressed with Bayesian approaches.
\cite{Robinson1978} proposes a way to estimate moments (with consistent and asymptotically normal estimators) that can be used as prior information. \cite{Liu1980} show that the prior on the parameters of the distribution of the random coefficients has progressively less influence as the number of units increases.
For a RCAP(1), they derive the theoretical posterior distribution of the coefficients in case of beta distributed AR coefficients, discuss how the model can be applied to seasonal data and how it can be extended to second order autoregression.
\cite{Li1983} propose empirical Bayes estimates for RCAP($p$), with a simpler implementation and no need for the prior distribution of the coefficients compared to the method of~\cite{Liu1980}, and show that the newly proposed approach outperforms LS estimation in most of the cases.
In fact, the limit distributions of empirical Bayes, Bayes and frequentist estimates are asymptotically equivalent~\citep{Kim1992}.
The independent multivariate student's t--inverse gamma prior is shown to lead to the best results among three priors in a genetic study, while on simulated data, the method performs well with all the tested priors~\citep{Safadi2011}.
Authors have also developed a Metropolis-within-Gibbs sampling algorithm and two algorithms based on MCMC (only one requiring stationarity)~\citep{Pai1994, Nandram1997}.
\cite{Nandram1997} state that their method performs better than VAR or BVAR, both on stationary and non-stationary time series, and analyze the same case study investigated by~\cite{Liu1980}.
Other recent publications use model-based approaches similar to the ones described in this section for clustering~\citep{Nascimento2012, Wang2012, Nascimento2016}.


\subsubsection{Autoregressive linear mixed models (ARLM)}\label{sec:ARLM}
When the random autoregressive and moving average coefficients are null almost surely, \eqref{eq:UAR} reduces to the autoregressive linear mixed ARLM($p$) model of order $p$ of~\cite{Funatogawa2007},
\begin{equation}
\YY{i}{t} = \sum_{k=1}^{p}\fY{k}{} \YY{i}{t-k} + \design{X}{i}{t} \fX + \design{Z}{i}{t} \rZ{i}{} + \ee{i}{t}. \label{eq:ARLM}
\end{equation}
Here the autoregressive coefficient is constant, and the heterogeneity is due to the random effects $\rZ{i}{}$. The assumptions on the random coefficients are these traditional in linear mixed models (LMM), $\rZ{i}{}$ and $\ee{i}{t}$ are assumed to be independent across units $i$, both iid normal with constant variances $\Sigma_c$ and $\Sigma_{e}$ respectively.
The authors introduce this model for the specific {\bf application} of fitting dose-response profiles in clinical trials showing initial sharp changes, decreasing rates of change, and finally approaching random patient-specific asymptotes.
\cite{Funatogawa2007} discuss the properties of the ARLM model, and via a simulation and case study they conclude that their model outperforms previous approaches for this specific application.
\cite{Funatogawa2008b} focus on the effect of drop-outs on the asymptotes' estimates and consider estimation in case of unequally spaced measurements in time~\citep{Funatogawa2012}.
The same authors also introduce the bivariate model and analyze the case in which the dose is based on previously observed responses~\citep{Funatogawa2008a, Funatogawa2012a}.
{\bf Estimation} is performed via ML, shown to be consistent, in some cases combined with a state-space representation to enable Kalman filter estimation~\citep{Funatogawa2008}.
\cite{Funatogawa2007} propose a reparametrization of~\eqref{eq:ARLM} that makes the connection with traditional LMM more evident, allows the extension to higher order autoregression, and the {\bf implementation in standard software} for LMM.
This method does not work when intermittent (i.e.,~followed by observed values) missing values are present.

\paragraph{Autoregressive panel data models (ARP)}\label{sec:ARP}
The autoregressive (or dynamic) panel data (ARP) model,
\begin{equation}
\YY{i}{t} = \sum_{k=1}^{p}\fY{k}{} \YY{i}{t-k} + \design{X}{i}{t} \fX + \rZ{0}{i} + \ww{i}{t} \label{eq:ARP}
\end{equation}
is a special case of~\eqref{eq:ARLM} with random intercept only, although no cross references exist between the two model families, 
applied in contexts with {$(n, 1,T)-$}data with small $T$.
This structure is particularly popular in econometric, economic and psychology {\bf applications}, where typically one has {$(n, 1,T)-$}data with small $T$. 
{\bf Estimation and identifiability} are discussed in two consecutive publications by~\cite{Anderson1981, Anderson1982}.
The authors also investigate the consistency of ML estimates and asymptotic properties as both $T$ and $n$ diverge to infinity, evaluating the influence of initial conditions.
\cite{Anderson1982} extend model~\eqref{eq:ARP} allowing for the inclusion of both time-invariant ({\itshape serial correlation model\/}) and time-varying ({\itshape state dependence model\/}) exogenous variables and show which parameters can be estimated.
Their work fits the general framework of~\cite{Macurdy1982}, structuring error models.
For {estimation}, the generalized method of moments (GMM) estimator has large finite sample bias and poor precision 
because the series are highly autoregressive and the number of observations is typically moderately small.
Various alternative approaches exist, often exploiting instrumental variables.
A comparative study illustrates how a modified version of the least squares dummy variable estimator (i.e.~including dummy variables to eliminate individual effects) can outperform GMM in terms of asymptotic variance, while achieving small bias~\citep{Kiviet1995}.
Furthermore, various modified versions of the GMM have been investigated, such as the Arellano-Bond estimator and the system GMM~\citep{Arellano1991, Ahn1995, Blundell1998, Blundell2000, Blundell2001, Bond2002}.
Few references study Bayesian inference in ARP models under either standard and non-Gaussian assumptions~\citep{Hirano2002, Juarez2010}.\\
The fundamental problem of {\bf testing} the presence of individual effects~\citep{Holtz1988, Arellano1991}, serial correlation~\citep{Wooldridge2002, Drukker2003} and unit roots~\citep{Levin1992} in dynamic panel data models is addressed by many authors to enable the choice of the most parsimonious structure and to test stationarity.

The number of publications studying (mostly economic) {\bf applications} of model~\eqref{eq:ARP} is extremely large.
More information on model selection and estimation methods can be found in comparative papers such as the ones from~\cite{Judson1999} and~\cite{Bond2002}.
The large interest in these models is reflected in the number of {\bf software packages} available for simulation and inference.
We mention for reference the plm package~\citep{Croissant2008}, the cquad package~\citep{Bartolucci2015} allowing dynamic binary panel data, the OrthoPanels package~\citep{OrthoPanels} using the orthogonal reparametrization approach, and the panelvar package~\citep{panelvar}.

\subsubsection{Time-series-cross-sectional models (TSCS)}\label{sec:TSCS}
Time-series-cross-sectional (TSCS) models of~\cite{Beck1995} constitute a niche independent of the rest of the literature.
The main reason for this is probably the type of data being analyzed, large in $n$ and $T$, and the asymptotics, consequently studied with respect to both dimensions.
The model is widely {\bf applied} in political economy and can be stated in the following state-space form
\begin{equation}
\begin{aligned}
\YY{i}{t} &=  \design{X}{i}{t} \fX +  \ww{i}{t}\\
\ww{i}{t} &= \sum_{j=1}^{r} \re{i}{j}{} \ww{i}{t-j}   +\ee{i}{t}. \label{eq:TSCS}
\end{aligned}
\end{equation}
where $e_{i,t}$ are independent identically distributed random variables with zero mean.
Note that~\eqref{eq:TSCS} is in fact a UAR model with $\rY{i}{}{}=0$ a.s., $\fY{k}{}=0\ \forall k$ and $\design{Z}{i}{t}=0\ \forall i,t$.
For {\bf estimation}~\cite{Beck1995} propose the so-called Parks method, warning that the generalized least squares (GLS) method underestimates variability in TSCS data unless $T\gg n$.
Feasible GLS (FGLS) is discussed as an alternative.
The authors conduct a Monte Carlo simulation study to show the importance of including individual autoregressive coefficients in small data-sets, and suggest the use of panel-corrected standard errors which take heterogeneous serial correlations into account.
\cite{Beck1998} extend the model to time series with a binary response variable and a large number of observed units.
The authors also alert researchers that taking correlation in time into account is crucial for proper inference, and show the consequences of neglecting correlations by looking at previously published studies where correlations were ignored~\citep{Beck1998, Beck2001, Beck2001a}.
\cite{Podesta2006} compares the goodness of fit of various models on the specific case study of Welfare State development.

\subsection{Bayesian vector autoregressive models (BVAR)}\label{BVAR}
Bayesian vector autoregressive (BVAR) models are the traditional vector autoregressive models, when the autoregressive coefficients are estimated via Bayesian approaches by setting a prior (i.e.~shrinkage methods). This, jointly with the likelihood of the data, returns a posterior distribution for the coefficients. Bayesian methods are shown to be appropriate in the estimation of large dynamic models~\citep{Banbura2010}.
A BVAR model of order $p$ is in fact a RARMA($p,0,0,0$) model, with $\design{X}{i}{t}=\design{Z}{i}{t}=0\ \forall i,t$, $\rforecast{i}{t}=0$ a.s.~and serially independent residuals. However, BVAR models do not fulfill our inclusion criteria since the randomness on the coefficients is imposed for estimation, while we consider models where the coefficients are random by assumption (inclusion criteria 2.).
We refer thus to other references for more details~\citep{Litterman1986, De2008, Banbura2010, Wozniak2016}.

\subsection{Hierarchical vector autoregressive models (HVAR)}\label{HVAR}
In recent years, {\bf studies} in psychology and behavioral science have started combining dynamic models for temporal behavior with hierarchical structures to account for both inter- and intra-individual variability.
These models address data where all dimensions ($n,\ m,$ and $T$) are large, and developed almost independently of the remaining literature. 
Since a definition does not exist and the models considered here vary also in terms of structure, we collect them under the general term hierarchical vector autoregressive (HVAR) models - highlighting the multivariate outcome and the hierarchical definition, without forcing a specific model structure.

This includes, for instance, the multilevel VAR model of~\cite{Bringmann2013}, that is in fact a RCAP~\eqref{eq:RCAP} suited for ($n,m,T$)-data. Given the field of application and the fact that this model constitutes the basis for a network approach, we treat it separately. It is also interesting to notice the analogy with BVAR models -- in the sense that the model equations are the same, but here the auto- and cross-regression coefficients are random for the model assumptions, and {\bf estimation} is based on pseudo-likelihood. The model of~\cite{Bringmann2013} defines subject-specific equations that enable comparisons across groups.

Another example in this model family is the hierarchical state space model of~\cite{Lodewyckx2011} (c.f.~\cite{Oravecz2011, Ranganathan2014} for the continuous-time version), a dynamic linear model with subject-specific parameters, that can be rewritten as
\begin{equation*}
\begin{aligned}
\YY{i}{t} &= \design{Z}{i}{t} \rZ{i}{t} +\ww{i}{t}\\
\ww{i}{t} &= \sum_{j=1}^{p}\re{i}{j}{} \ww{i}{t-j} +\ee{i}{t},
\end{aligned}
\end{equation*}
with the notation introduced in Section~\ref{sec:RAR}. It is in fact a RARMA($0,0,p,0$) model with $\rforecast{i}{t}=0$ a.s.~and an additive structure for $\ee{i}{t}=z_i\design{X}{t}{i}+v_t$. The authors take a Bayesian approach for the {\bf estimation}.
Furthermore, hierarchical Bayesian models have received particular attention and they are shown to be well suited for {\bf studies} in psychology~\citep{Shiffrin2008, Lee2011, Ranganathan2014}. In fact, also {\bf software packages} exist (e.g.~BVAR~\citep{BVAR}).
\cite{Adolf2014} have also introduced a {\bf test} to check whether inter- and intra-individual model structures are equivalent.

\section{Models for heterogeneity in time and across units}\label{sec:TimeUnitHeterogen}
In this section we describe models that can accommodate heterogeneity both in time and across units.
The availability and dimension of data-sets have increased significantly in the recent years, and require methods capable of addressing multiple properties of the data simultaneously.
In this direction, the random coefficient panel model has been recently developed by~\citep{Horvath2016}.
Typical example {\bf applications} include macroeconomic and financial studies, but also the modern challenge of analysing multiple channels of electroencephalography (EEG) data and speech signals. 
Also the dynamic factor model of~\cite{Prado2001} aims at capturing multiple dynamics present in the data and is particularly suited for the analysis of EEG data.

\subsection{Random coefficient panel models (RCP)}\label{sec:RCP}
With the notation of Section~\ref{sec:RAR}, a random coefficient autoregressive panel (RCP) model~\citep{Horvath2016} is defined as
\begin{equation}
\begin{aligned}
\YY{i}{t} &= \sum_{k=1}^{p}\fY{k}{} \YY{i}{t-k} + \design{X}{i}{t} \fX + \design{Z}{i}{t} \rZ{i}{} + \ww{i}{t},\\
\ww{i}{t} &= \sum_{j=1}^{s} \rY{i}{t}{j} \YY{i}{t-j} +\ee{i}{t}, \label{eq:RCP}
\end{aligned}
\end{equation}
which is in fact a RARMA($p,0,0,p$)~\eqref{eq:RARMA} model with $\rforecast{i}{t}=0$ almost surely, the autoregressive coefficients independent of the residuals and all the stochastic terms serially independent, and independent across units. 
Note that the nomenclature could be confused with that of autoregressive panel data models, treated in Section~\ref{sec:ARP}, but we make the choice here not to introduce additional terms.
Model~\eqref{eq:RCP} is suited for $(n,1,T)-$data, and is capable of addressing heterogeneity both in time and across units, with subject-specific stochastic autoregressive coefficients and the following factor structure for the error term
\begin{equation}
\ee{i}{t} = z_{i,t} + \chi_i v_t.\label{eq:mult_structure_e}
\end{equation}
Here, the term $v_t$ has zero mean and unit variance, while $\chi_i$ is independent across units and independent of the other random terms.
Random coefficient panel models include both RCA and RCAP models as particular cases for some filtration, $\design{X}{i}{t}=\design{Z}{i}{t}=0,\ \forall i,t$ and no structure is imposed on $\ee{i}{t}$.

In their publication, \cite{Horvath2016} show that the unit root problem exists only in case $T\longrightarrow \infty$, and prove that the WLS {\bf estimator} is asymptotically normal.
Furthermore they illustrate that this estimator performs well also for relatively small panel data (both in terms of $T$ and $n$) via a simulation study. A macroeconomic and a financial applications, motivating the study, are included in the publication.

\subsection{Dynamic factor models (DFM)}\label{sec:DFM}
Bayesian TVAR models (c.f.~Section~\ref{sec:TVAR}) are suited for single non-stationary time series. These models have been extended to pool information from multiple non-stationary time series, and to understand the cross and the spatio-temporal relationships.
These dynamics are of interest, for example, in the {\bf analysis} of multiple channels of EEG data and speech signals.
In dynamic factor models (DFM) the state follows a dynamic linear or TVAR model~\citep{West2006}, while the observation equation can assume various forms. Again, as for the TVAR model, multiple assumptions are possible. One example is the regression with fixed factor weights to adjust the effect of the latent variable for different channels~\citep{Prado1997, Prado2001},
\begin{equation}
\begin{aligned}
\YY{i}{t} &= \chi_i x_t + \nu_{i,t},\\
x_t &= \sum_{j=1}^p \phi_{t,j} x_{t-j} + \eta_t,\\
\phi_t &= \phi_{t-1} + w_t.
\end{aligned}
\end{equation}
It is in fact a RARMA($0,0,p,p$) model with $\rY{i}{t}{}=-\re{i}{t}{}=\phi_t$, assumed to follow a random walk, and a multiplicative structure for the residuals ($\ee{i}{t}=\chi_i \eta_t$) similar to~\eqref{eq:mult_structure_e} of~\cite{Horvath2016}. In this case, however, $\chi_i$'s are fixed factors, and thus the inclusion of this class into the models for the heterogeneity across unit and time is questionable.
It is also interesting to note the analogy with HVAR models, but here the autoregressive coefficients are random in time while in HVAR they are random samples from the population.


\section{Discussion}\label{sec:Discussion}
In this paper we presented a structured overview of models with autoregressive structure and random coefficients.
The existing literature, broad and disconnected, is structured and made accessible also to the reader who is new to the field.

First we provided a concise summary of the terminology used for data-sets and models.
Then we introduced the RARMA structure to provide a unified language, and to show the hierarchy existing among models in a formal way.
Through a mathematical approach, we stated the simplifying assumptions necessary to get the nested models from the more complex ones. This way, also similarities and differences are shown explicitly.
The exposition is supported by a graph and a table, that help visualizing the hierarchy and getting the variety of existing models at a glance.

Since the literature is fragmented, bringing together traditionally independent fields will boost the research in each field, exploiting the achievements of others.
The present work supports this evolution, by also providing an overview of properties, existing estimation methods and tests that can be exploited and expanded in different directions to adapt to various types of data. This will also avoid replication of results, which is quite common in this context.
In each section, where available, software packages or code made available with publications are listed. However, it is apparent that while the literature about these models is extremely broad, it is not supported by a corresponding availability of implementations, further limiting the reproducibility and the possibility of advances in the field.
Some subfields like (G)ARCH and panel data models are well developed also in terms of software, but future contributions are needed in other fields like (G)RCA.



We limited this review to autoregressive models with random parameters for discrete-time data, but extensions in multiple directions are possible.
For example, it  would be interesting to extend this work to other autoregressive models, e.g.~models with deterministically time-varying parameters. 
It would also be valuable to extend the overview to continuous-time models.
Further research should focus furthermore on the analysis of the newly introduced RARMA model, checking identifiability and proposing suitable estimation methods.
Hopefully the results will move towards open source implementations, differently from the current literature -- which is vast and not easily reproducible. 
%
%


\section*{Acknowledgements}
This research was performed within the framework of the strategic joint research program on Data Science between TU/e and Philips Electronics Nederland B.V.


\section*{Conflict of interests}
No potential conflict of interest was reported by the authors.

\section*{Data availability statement}
Data sharing is not applicable to this article as no new data were created or analyzed in this study.

\bibliographystyle{myplainnat}
\bibliography{References_new}

\end{document}